# Single Crystal Growth and Spin Polarization Measurements of Diluted Magnetic Semiconductor (BaK)(ZnMn)$_2$As$_2$

G. Q. Zhao[1,2], C. J. Lin[1], Z. Deng[1], G. X. Gu[1], S. Yu[1], X. C. Wang[1], Z. Z. Gong[3], Y. J. Uemura[3], Y. Q. Li[1,2], C. Q. Jin[1,2]

[1] Institute of Physics, Chinese Academy of Sciences; Collaborative Innovation Center of Quantum Matter, Beijing 100190, Chin
[2] School of Physics, University of Chinese Academy of Sciences, Beijing, 100190,China
[3] Department of Physics, Columbia University, New York, NY 10027, USA

**ABSTRACT**

Recently a new type diluted magnetic semiconductor (Ba,K)(Zn,Mn)$_2$As$_2$ (BZA) with high Cure temperature ($T$c) was discovered showing independent spin and charge doping mechanism. This makes BZA a promising material for spintronics devices. Here we report for the first time the successful growth of BZA single crystal. An Andreev reflection junction that can be used to evaluate spin polarization was fabricated based on the BZA single crystal, a 66% spin polarization of the BZA single crystal was hence obtained by Andreev reflection spectroscopy analysis.

**Introduction**

Diluted magnetic semiconductors (DMSs) have triggered extensive research due to their fantastic physical properties and applications for spintronic devices since the discovery of (Ga,Mn)As film by H. Ohno in 1990s.[1-7] In these Ⅲ-V DMSs, such as (Ga,Mn)As and (In,Mn)As, substitution of divalent Mn atoms into trivalent Ga (or In) sites leads to severely limited chemical solubility, resulting in metastable specimens only available as epitaxial thin films.[2] The hetero-valence substitution, which simultaneously dopes hole carriers and spin, makes it difficult to individually control charge and spin concentrations for flexible tuning of quantum freedom of DMS. To solve these problems, several new types of DMSs with independent spin or charge doping were synthesized, e.g. the "111" type Li(Zn,Mn)As、"122" type (Ba,K)(Zn,Mn)$_2$As$_2$ and "1111" type (La,Ca)(Zn,Mn)SbO, which are named by the chemical ratio of their parent phases.[8-23]

Among these new DMSs, the ThCr$_2$Si$_2$ type (Ba,K)(Zn,Mn)$_2$As$_2$ has a Curie temperature ($T$c) up to 230 K which marks the current reliable record $T$c for DMSs where ferromagnetism is mediated by carriers.[12-13] The (Ba,K)(Zn,Mn)$_2$As$_2$ is believed to be one of milestone materials for the research of DMSs.[24] A robust nearest-neighbor ferromagnetic correlation that exists above the ferromagnetic ordering temperature suggested potential to realize even higher $T$c in further study.[25] Angle-resolved photoemission spectroscopy (ARPES) measurements showed clear impurity band of doping Mn well below the Fermi energy.[26-27] Besides, the excellent match of lattice parameters (within 5% mismatch) among "122" type DMS (Ba,K)(Zn,Mn)$_2$As$_2$, "122" iron-based superconductor (Ba,K)Fe$_2$As$_2$, and antiferromagnetic BaMn$_2$As$_2$ is promising for fabricating heterojunctions with different types of ordering.[13] Thus the BZA will provide a unique opportunity to elucidate the intrinsic physics in DMSs and the physically transparent description of them may also be general and applicable to other

DMS materials.[24,28,29] For both fundamental understanding and potential applications on spintronics devices, spin polarization (*P*) of BZA from direct measurement is an important parameter. The Andreev reflection (AR) technique has been applied to measure the spin polarization rate of prototypical III-V based DMS, e.g. 85% for (Ga,Mn)As,[30] 57 ± 5% for (Ga,Mn)Sb,[31] and 72% for (In,Mn)As.[32] Single crystals with various K and Mn doping levels have been grown and the *T*c of crystals are controlled by K and Mn concentrations, i.e. carrier and spin density, respectively. As an initial attempt, the crystal of $(Ba_{0.904}K_{0.096})(Zn_{0.805}Mn_{0.195})_2As_2$ that shows good shape and size was selected to fabricate the Andreev reflection junction. Here we report the basic properties of $(Ba_{0.904}K_{0.096})(Zn_{0.805}Mn_{0.195})_2As_2$ single crystal, along with the degree of spin polarization obtained from the crystal based Andreev reflection spectroscopy.

**Results and Discussion**

**Chemical Composition and Crystal structure.** The chemical compositions and morphology of the single crystal were investigated by using energy dispersive X-ray analysis (EDX) and Inductively Coupled Plasma (ICP) mass spectrometry. The real atom ratio, $(Ba_{0.904}K_{0.096})(Zn_{0.805}Mn_{0.195})_2As_2$, was determined by ICP. We also used the EDX to analyze the real atom ratio and the doping homogeneity. The results are consistent with the ICP results. Fig. 1 shows the obtained BZA crystals with typical size of 3×3 mm$^2$. The X-ray diffraction patterns of the obtained crystals only shows the (002n) peaks of the $(Ba,K)(Zn,Mn)_2As_2$ structure as shown in Fig. 1. The unit cell constants are calculated to be c=13.4658(6) Å which are consistence with previous reports.[13] To further confirm the phase, the single crystals were ground to conduct powder X-ray diffraction. The obtained pattern fits very well with $ThCr_2Si_2$ structure.

**Magnetic properties.** The DC magnetic susceptibility of the BZA single crystal was characterized using a superconducting quantum interference device magnetometer (Quantum Design) in both zero-field-cooling (ZFC) and field-cooling (FC) mode. Both of the in-plane magnetization $M$ versus $T$ data ($M_{ab}(T)$) and the $H // c$ axis $M_c(T)$ at $H = 500Oe$, shown in Fig. 2(a), exhibit clear ferromagnetic enhancements at ∼50 K. A precise determination of $T_c$ can be done by the critical exponent analysis, which requires a fine measurement of the M-H data in a sufficient small temperature interval over a large temperature region. However, it is not the subject at present work. Ferromagnetism is also evident from the $M(H)$ plots shown in inset of Fig. 2(b) with a saturation moment $M_{sat}$ of about 0.5 and 0.3($\pm 0.03$) $\mu_B$/Mn in $M_c(H)$ and $M_{ab}(H)$, respectively. The $M_{sat}$ is defined as high-field $M(H)$ data at 2 K after subtracting this small T-linear component.[6] As discussed in our previous paper about the polycrystalline samples, the antiferromagnetic coupling of Mn in the nearest neighbor Zn sites can reduce the saturation moment and also cause a linear component on M(H) curves simultaneously.[18] The small T-linear component of current single crystal is calculated as 0.059$\mu_B$/T along c-axis and 0.057$\mu_B$/T along ab-plane. The coercive force, $H_c^c$, in $M_c(H)$ is about 5300 Oe and $H_c^{ab}$ in $M_{ab}(H)$ is about 1200 Oe. The $H_c^c$ and the $H_c^{ab}$ become smaller when temperature rises while $M_{sat}$s along c axis are always larger than that in ab-plane at any temperature from 10K to 105K as shown in Fig. 2(c) and Fig. 2(d), respectively. The crystals show clear anisotropic behavior with easy axis along c from the M(T) and M(H) measurements.

**Electrical transport properties.** Fig.3 (a) shows the temperature dependence of resistivity with electrical current in ab plane ($\rho_{ab}(T)$). It shows an increase of resistivity as decreasing temperature due to the semiconductor behavior and the localization effect.[33] The magnetoresistance (MR) and the Hall effect measurements were performed with the electrical current in ab plane from 2K to 130K and with the magnetic field parallel to c axis up to 14 T.

Fig.3 (b) shows change of (MR-$R_{xx}$) at several selected temperatures from 2 K to 130 K and Fig.3(c) shows the corresponding Hall resistance $R_{xy}$. The negative slope in Hall resistance at high magnetic field indicated p type carrier, which is consistent with the substitution of monovalent K into divalent Ba. The salient features of $R_{xx}$ and $R_{xy}$ are the gradually emerging of hysteresis at temperatures below 10 K, from which a coercive field $H_c^c$ around 5300 Oe can be clearly identified at 2K, which is in good accordance with the magnetization measurement shown in Fig.2 (b). From the transport measurements, we observe non-linear Hall resistance at low magnetic field up to 70 K. Above 70 K, the Hall resistance becomes linear, which according to the anomalous Hall effect suggests the linear field dependence of the magnetization was obeyed. This indicates there is no spin correlation effect. However, this temperature should not necessary be the same as the ferromagnetic transition (long rang order) temperature if there exist a region with short-range spin correlation e.g. in GaMnAs[1], also recently observed in (Ba,K)(Zn,Mn)$_2$As$_2$[25]. Therefore, these two temperatures, 50 K and 70 K represent two critical points with different type of spin correlation for long range ordering & short rang fluctuations, respectively. In addition, we also noticed the "overlap" in (MR-$R_{xx}$) between 50K and 60K which is near the $T$c as shown in Fig.3 (b). It results from the suddenly reduction of MR above $T$c (here 60K). This phenomena was also observed in (Ga,Mn)As.[1,34]

In order to determine the carrier density for BZA, we take a more quantitative analysis to the hall resistance. Generally, in the ferromagnetic statue of a DMS material, the scattering from the magnetic ions causes the asymmetric accumulation of carriers in the transverse direction relative to the electric current, giving an additional contribution to the normal Hall effect which is called as anomalous Hall effect (AHE).[33] The Hall resistance therefore can be phenomenological expressed with two terms as

$$R_{xy} = R_0 B + R_s M(B), \qquad (1)$$

where $R_0$ is the ordinary Hall coefficient, $R_s$ is the anomalous Hall coefficient and M is the magnetization moment. As we mentioned in above, a small paramagnetic background was found in field dependent magnetization measurements at low temperatures in this material, it was until the magnetic field reached to ~ 11-14T, the magnetization then got saturated. In Fig.3(c), it is noticeable that $R_{xy}$ at these high field regions are almost straight lines, this implies the dominance of one type carrier near the Fermi surface that takes response to the magnetic field. Therefore, a single band model of Equ.1 is justified for Hall effect analysis in BZA. Furthermore, since the magnetization saturated at high magnetic field, the anomalous hall resistance $R_s M(B)$ becomes field independent, therefore by derivative of $R_{xy}$ in Equ,1 with magnetic field B, we can deduce the ordinary Hall coefficient $R_0$ which is simply equivalent to the high field slope of $R_{xy}$ in Fig.3(c). Then the hole carrier density $n_p$ for every temperature can be deduced by the relation $n_p = 1/e|R_0|$. The $n_p$ vs T is plot in Fig.3(d). As we can find the carrier density increases monotonically from $2.82 \times 10^{20}$ at 2 K to $4.80 \times 10^{20}$ cm$^{-3}$ at 130 K. Much similar to a semiconducting behavior, the observed increase of the carrier density when raising temperatures may due to the enhanced thermal associated excitation of carriers from the impurity bands to the conduction band.

**Spin polarization.** The spin polarization rate is one of key parameters of a DMS material for direct fundamental and applied relevance. *P* of various traditional DMS materials have been determined by analysis of Andreev reflection spectroscopy. Similarly, we use Andreev reflection spectroscopy to directly probe the electron spin polarization in single crystal BZA. This method has been successfully applied to measure the spin polarization in (Ga,Mn)As[30], (Ga,Mn)Sb[31], (In,Mn)As[32] along with other ferromagnetic materials, such as(La,Sr)MnO$_3$[35], CrO$_2$[36], EuS[37] and HgCr$_2$Se$_4$[38]. The inset of Fig.4 shows a schematic view of the BZA/Pb junction. The typical junction area is around 100*100 μm². The differential conductance, defined as G($V$)=dI($V$)/d$V$ was measured as a function of the dc bias voltage V crossing the junction by using phase-sensitive lock-in techniques. The amplitude of the ac modulation outputted from the lock-in amplifier was kept around 20 nA that is sufficiently small in order to avoid spurious artificial effects. The normalization for the differential conductance G to $G_0$ was carried out with $G_0$ measured in magnetic field of 0.25 T. In Fig.4, we present the temperature dependence of $G/G_0$ from 1.7K to 35K, for which a dramatic drops appears at T=7.2K. This temperature is exactly corresponding to the superconducting phase transition of Pb and the dropping of $G/G_0$ confirmed the Andreev reflection process has been taken place at interface between BZA single crystal and the superconducting Pb film. Furthermore, from the plot of $G/G_0$ vs. dc bias V in Fig.5 with several temperatures from 1.7 K to 7 K, we observed a suppression of the Andreev reflection spectra inside the superconducting gap which can be attributed to the ferromagnetism originated spin-imbalanced density of states around the Fermi level in BZA that partially inhibits the formation of Cooper pairs and their tunneling into the superconductor.

To be more quantitatively, we use the Modified Blonder-Tinkham-Klapwijk (BTK) theory[22] to describe the spectra by fitting with three parameters: superconducting gap Δ, interface barrier Z, and the spin polarization P. The fitting curves have been presented in Fig.5 together with the experimental data as used for a comparison. From the figure, we can see that well fittings to all the data in the whole temperature ranges have been achieved and the extracted fitting parameters were presented in Fig.5. Since in the fitting results for different temperatures, we only found that the parameter of superconducting gap Δ shown a decreasing behavior in elevated temperatures, this therefore strongly suggests the parameter P and Z have been well determined in the fitting process and the temperature dependence of Δ, Z and P are all behaved in expects from the BTK theory. Moreover, we notice that the fittings have resulted a small $Z$ value ($Z=0.38\ll1$), which implies a clean and transparent interface between BZA crystal and Pb film has been achieved in our junctions. Because a small $Z$ has been regarded as a crucial required in spectra analysis as it directly warrants the reliability and accuracy for extracting the spin polarization in the fitting process, therefore our result on $Z$ further convinced the determination of the spin polarization $P$ in our Andreev reflection spectra analysis, which gives $P=66\pm1\%$ for BZA single crystal.

**Conclusions**

In summary, we have successfully grow the single crystal of $(Ba,K)(Zn,Mn)_2As_2$ system for the first time. The crystal shows a ferromagnetic transition with easy magnetization axis along c axis. The carrier density is deduced from the anomalous Hall effect to be from $2.82\times10^{20}$ to $4.80\times10^{20}$ cm$^{-3}$ as the temperature increases from 2K to 130K. More significantly, the Andreev reflection junction from the selected large size single crystal was fabricated to testify spin polarization degree of $(Ba,K)(Zn,Mn)_2As_2$ and 66% spin polarization is reached. The success on Andreev reflection junction opens up a solid route to further fabricate multilayer junctions based on $(Ba,K)(Zn,Mn)_2As_2$ DMS.

**Method**

Single crystal $(Ba_{0.904}K_{0.096})(Zn_{0.805}Mn_{0.195})_2As_2$ were grown by flux technique. Firstly, the precursor materials of (Zn,Mn)As mixture were prepared with high-purity Zn, Mn and As in a sealed tube. The samples were heated at 750 °C and held for several hours before the temperature was decreased to room temperature. Then the mixtures of precursors with high-purity Ba and high-purity K in appropriate molar ratio were loaded into niobium tube with argon under 1atm pressure before sealed into a quartz tube. The process was handled in a glove box with high-purity argon to protect the materials from reaction with air or water. The quartz tube was heated at 1200 °C and held for several hours before the furnace was cooled down to room temperature at a rate of 3 °C/h. The recovered samples were characterized by X-ray powder diffraction with a Philips X'pert diffractometer using Cu$K\alpha$ radiation. Real compositions were determined by using energy dispersive analysis (EDAX) on a commercial Scanning Electron Microscope (SEM) and Inductively Coupled Plasma (ICP) mass spectrometry. The DC magnetic properties were examined by using Superconductivity Quantum Interference Device (SQUID, Quantum design), and transport properties and Andreev reflection junction were examined by Physical Property Measurement System (PPMS, Quantum design) with lock-in techniques. During the transport experiments, the single crystals were cleaved to get a clean fresh surface in order to get a good ohmic contact. A standard four points method was employed to eliminate contact resistances with a center electrodes pad of 0.5mm×0.5mm by using sliver paint as electric contact while gold wire as electric leads. A current of 50 μA was used during all of the transport measurements.

**Acknowledgments**

Works at IOPCAS are supported by NSF & MOST of China through Research Projects, as well as by CAS External Cooperation Program of BIC (112111KYS820150017).The work at Columbia was supported by NSF DMR-1436095 (DMREF).

**Author contributions statement**

C. Q. J. conceived the experiments. G. Q. Z. and C. J. L. conducted the experiments. G. Q. Z., C. J. L., Y. Q. L. and Z. D. performed the data analysis. The obtained results were discussed with, G. X. G., S. Y., X. C. W., G. Q. Z., C. J. L., Z. D. and C.Q. J. wrote the manuscript. All authors reviewed the manuscript.

**Competing financial interests:** The authors declare no competing financial interests

# Figure Captions

**Figure 1.** The X-ray diffraction patterns of $(Ba_{0.904}K_{0.096})(Zn_{0.805}Mn_{0.195})_2As_2$ collected at room temperature. The inset show the crystal structure (right) and its photograph (left).

**Figure 2.** Magnetic properties of $(Ba_{0.904}K_{0.096})(Zn_{0.805}Mn_{0.195})_2As_2$ and its anisotropy. (a) DC magnetization measured along c axis and in ab-plane with ZFC and FC mode under external field $H$=500 Oe. (b) The hysteresis curves M(H) measured at 2K in deferent axis to exhibit magnetic anisotropy. (c)&(d) The hysteresis curves M(H) measured at selected temperatures from 10K to 105K in c axis and in ab-plane.

**Figure 3.** Transport properties of $(Ba_{0.904}K_{0.096})(Zn_{0.805}Mn_{0.195})_2As_2$ single crystal. (a) The temperature dependence of the resistivity with current in ab-plane. (b)& (c) The magnetoresistance $R_{xx}$ and the anomalous Hall effect $R_{xy}$ at several selected temperatures from 2 K to 130 K are presented. (d) The temperature dependence of the carrier density calculated based on $R_{xx}$ and $R_{xy}$ are shown.

**Figure 4.** Sketch of the $(Ba_{0.904}K_{0.096})(Zn_{0.805}Mn_{0.195})_2As_2$/Pb junctions used for Andreev reflection spectroscopy. The inset is the normalization for the differential conductance $G/G_0$.

**Figure 5.** Normalized differential conductance $G/G_0$ spectra (the red dot) and their fits to the modified BTK theory (the blue line) at selected temperature from 1.7K to 7.0K.

**Figures**

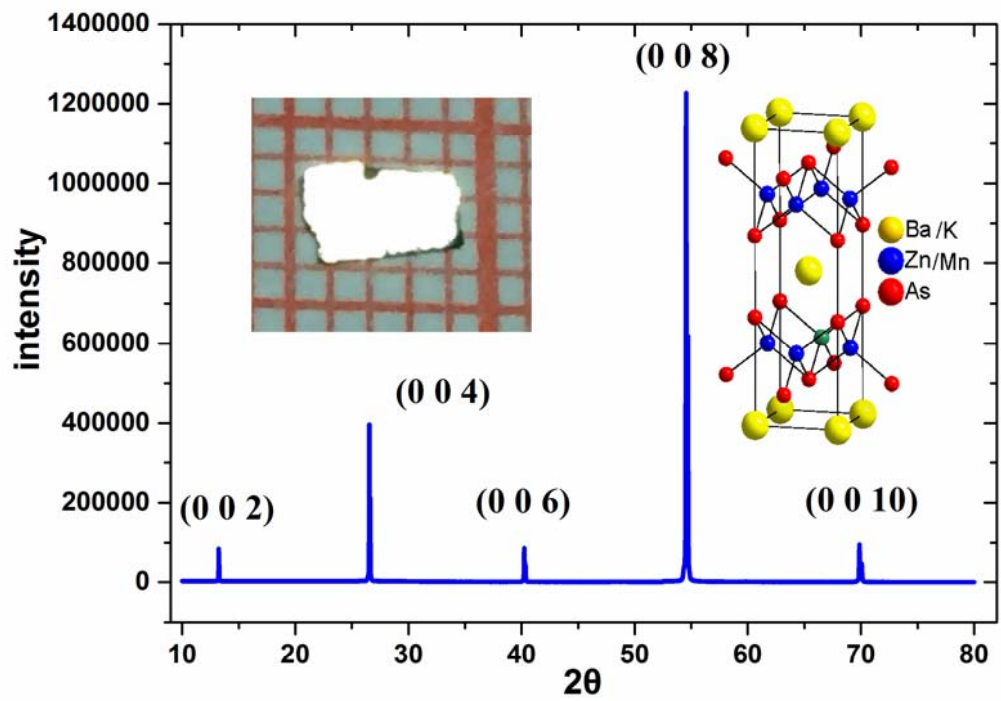

Figure 1

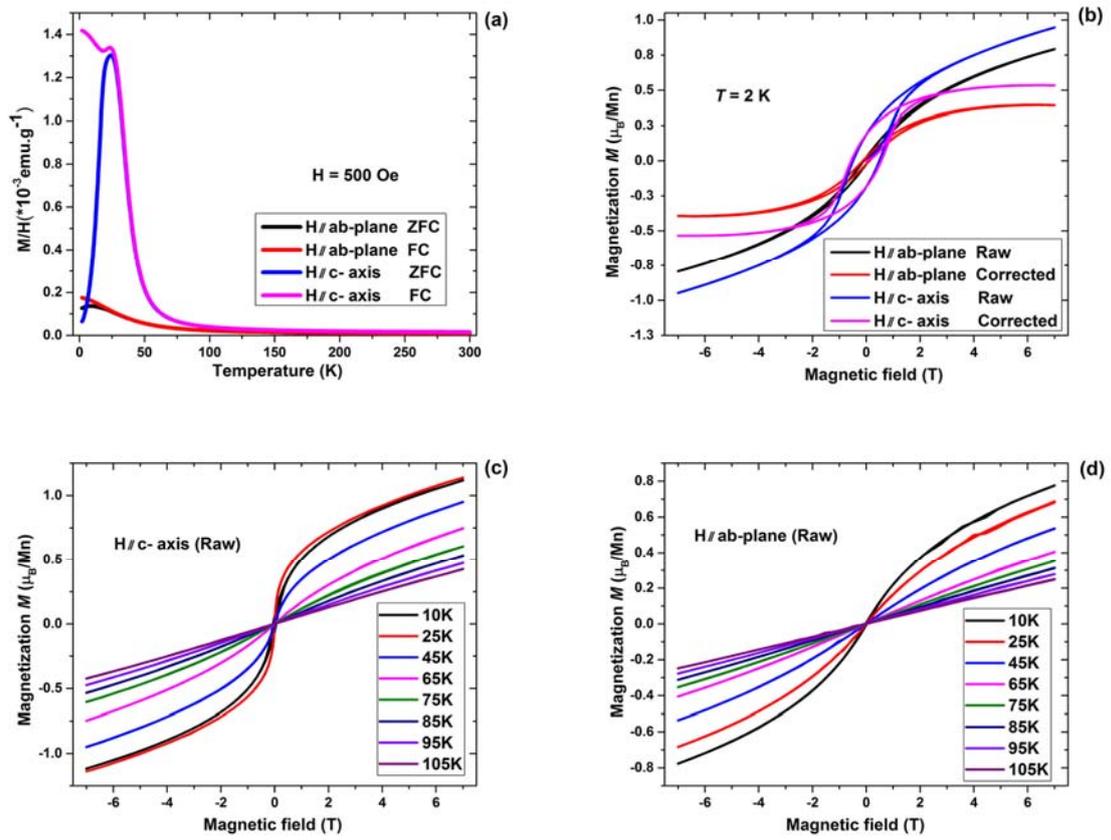

**Figure 2**

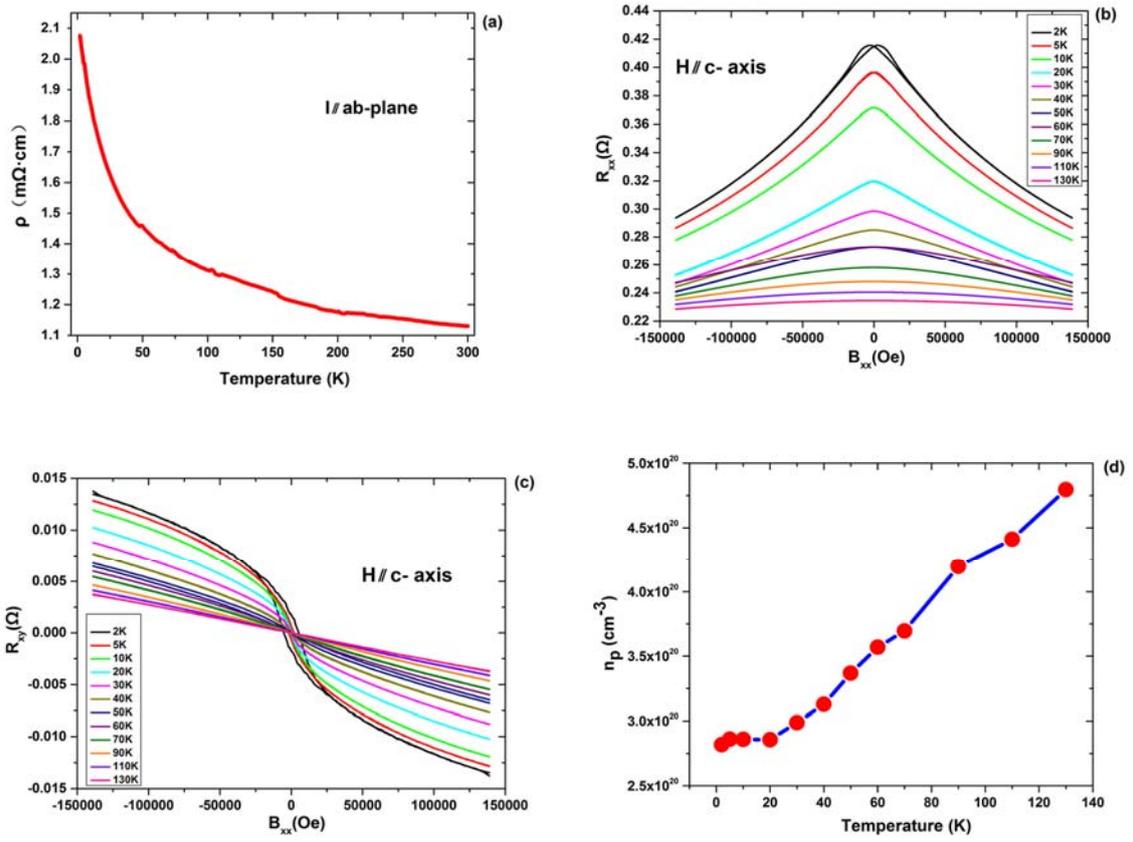

**Figure 3**

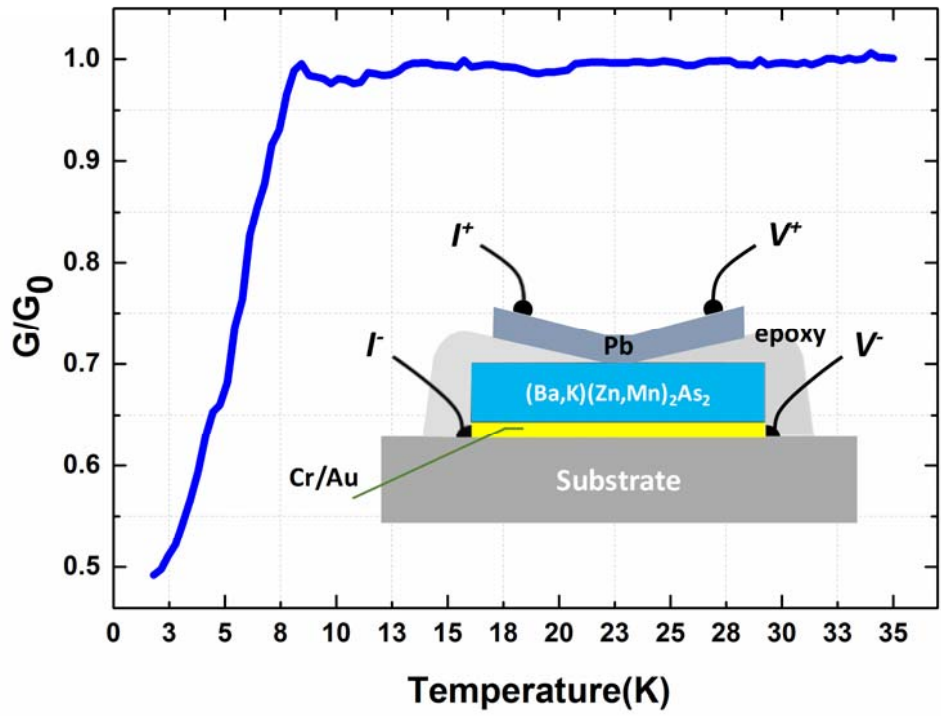

**Figure 4**

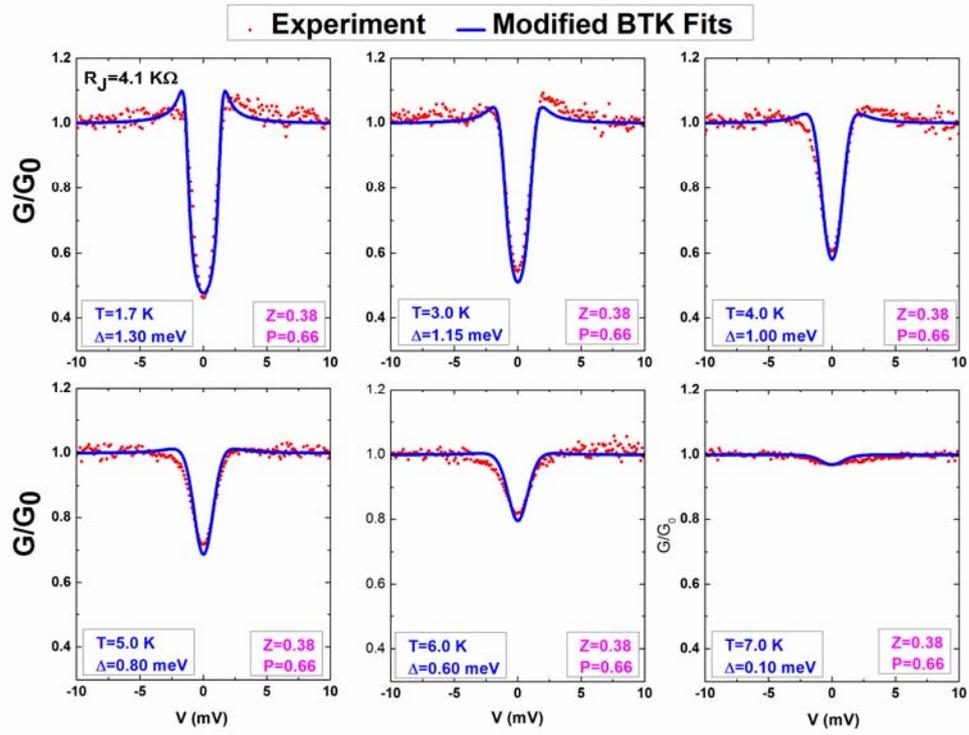

**Figure 5**